\documentclass[preprint,12pt]{aastex}
\usepackage{epsfig, amsmath, amsfonts}

\newcommand\bb[1] {   \mbox{\boldmath{$#1$}}  }

\newcommand\del{\bb{\nabla}}
\newcommand\bcdot{\bb{\cdot}}

\newcommand{\dd}[2]{\frac{{\rm d} #1}{{\rm d} #2}}

\def\dd{\partial}

\def\beq{ \begin{equation} }
\def\eeq{ \end{equation} }
\def\spose#1{\hbox to 0pt{#1\hss}}  
\def\ltsim{\mathrel{\spose{\lower.5ex\hbox{$\mathchar"218$}}
\raise.4ex\hbox{$\mathchar"13C$}}}
\def\gtsim{\mathrel{\spose{\lower.5ex\hbox{$\mathchar"218$}}
\raise.4ex\hbox{$>$}}}

\slugcomment{}

\begin{document}

\title{\bf\LARGE On Differential Rotation and Convection in the Sun}                 
\author{ Steven A. Balbus\altaffilmark{1,2}, 
Julius Bonart\altaffilmark{1},
Henrik N. Latter \altaffilmark{1},
Nigel O. Weiss\altaffilmark{1,3}}

\altaffiltext{1}{Laboratoire de Radioastronomie, \'Ecole Normale
Sup\'erieure, 24 rue Lhomond, 75231 Paris CEDEX 05, France
  \texttt{steven.balbus@lra.ens.fr}}

\altaffiltext{2}{Adjunct Professor, Department of Astronomy, University of Virginia,
Charlottesville VA 22903, USA}

\altaffiltext{3}{DAMTP, Centre for Mathematical Sciences, Wilberforce Road, 
 Cambridge CB3 0WA, UK}

\begin{abstract}
We show that the differential rotation profile of the solar convection
zone, apart from inner and outer boundary layers, can be reproduced
with great accuracy if the isorotation contours correspond to
characteristics of the thermal wind equation.  This requires that
there be a formal quantitative relationship involving the entropy
and the angular velocity.  Earlier work has suggested that this could
arise from magnetohydrodynamic stability constraints; here we argue
that purely hydrodynamical processes could also lead to such a result.
Of special importance to the hydrodynamical solution is the fact that
the thermal wind equation is insensitive to radial entropy gradients.
This allows a much more general class of solutions to fit the solar
isorotation contours, beyond just those in which the entropy itself
must be a function of the angular velocity.  In particular, for this
expanded class, the thermal wind solution of the solar
rotation profile remains valid even when large radial entropy gradients
are present.  A clear and explicit example of this class of solution
appears to be present in published numerical simulations of the solar
convective zone.  Though hydrodynamical in character, the theory is
not sensitive to the presence of weak magnetic fields.  Thus, the
identification of solar isorotation contours with the characteristics
of the thermal wind equation appears to be robust, accommodating, but
by no means requiring, magnetic field dynamics.

\end{abstract}


\maketitle


\section{Introduction}

In a recent paper, Balbus (2009, hereafter B09) argued that the shape
of the isorotation contours in the solar convection zone (SCZ) may be
understood as a consequence of a dominant thermal wind balance in the
vorticity equation, together with the near coincidence of surfaces of
constant specific entropy $S$ (isentropes) and surfaces of constant
angular velocity $\Omega$ (isotachs).  With $S$ a function of
$\Omega$, the thermal wind equation (TWE) may be solved analytically and
exactly; the isorotation contours then explicitly correspond
to the characteristics of the TWE.  B09 found that the agreement between
the characteristics and the observed $\Omega$ profile is very good,
even under the simplest of assumptions.  In this paper, we take a more
systematic approach, and find that the agreement is truly remarkable
(cf. figure [1]).   The quality of the fit is strong evidence not only
that thermal wind balance holds in the bulk of the solar convection zone
(as noted previously in many numerical simulations), but that there is 
also
some functional relationship between entropy and angular rotation
that needs to be understood.  

The hope, of course, is that this is telling us something important about
the physics of the solar convection zone.  B09 puts
forth the case that the SCZ is in a state of marginal dynamic instability
not only against convective instabilities, but against magneto-baroclinic
instabilities.  To understand why the distinction is important, recall
that a weak magnetic field in a differentially rotating gas is a catalyst
for destabilization.  This effect is well known to the accretion disk
community in the form of the magnetorotational instability (MRI), which
renders fully turbulent what would otherwise be hydrodynamically stable
Keplerian disks.  B09 argues that in its magnetobaroclinic guise, this
weak field instability can destabilize what would be hydrodynamically
stable stratified configurations of the SCZ, driving the system to a
condition of marginal instability.  Such a state is reached when constant
entropy surfaces and constant angular velocity surfaces nearly coincide,
leading to an $S=f(\Omega^2)$ relationship.

In this paper, we examine another possible explanation for the striking
fit evidenced by figure [1] in \S 2.  We are motivated, in part, by the
fact that some purely hydrodynamic calculations are able to reproduce
rather well certain noncylindrical features of the solar rotation profile
(Miesch, Brun, \& Toomre 2006, hereafter MBT06; Miesch et al. 2008;
Miesch \& Toomre 2009).  Those runs show that, provided there are
sufficiently large latitudinal entropy gradients at the base of the SCZ,
a very plausible fit to the solar isorotational contours can be found
throughout the bulk of this region.  In this current paper, we will argue
that there may well be a generic hydrodynamical mechanism that explains
both the observed helioseismology results and the numerical simulations.

An outline of our paper is as follows.  In \S 2, we present an improved
fit between an explicit solution of the thermal wind equation and the
helioseismology data.  We also introduce, following MBT06, the concept
of ``residual entropy'': the average entropy profile remaining after an
underlying radial profile has been removed.  Arguments are presented
which suggest that convecting fluid elements will tend to move in
surfaces of constant residual entropy.  In \S 3, we further argue that
the same fluid elements will also tend to move in surfaces of constant
angular velocity, and that these surfaces must therefore coincide with
those of residual entropy.  The coincidence of surfaces of residual
entropy and angular velocity is necessary in our approach to obtain
a solution of the thermal wind equation.  In \S 4 we conclude with an
evaluation and discussion of the relative merits of the hydrodynamical
and magnetohydrodynamical approaches.

\section{A simple model for the solar isorotation contours}
\subsection{Preliminaries}

\subsubsection {Coordinates and notation}

Let $(R, \phi, z)$ be a standard cylindrical coordinate system, and $(r,
\theta, \phi)$ a standard spherical coordinate system.  
We consider an equilibrium flow
in a state of azimuthal rotation in which the angular velocity $\Omega$
is assumed to be independent of $\phi$, but may depend upon $R$ and $z$.
The background entropy profile $S$ is also a function of $R$ and $z$.
Our notation for the other fluid variables is standard: $\bb{v}$ is
the velocity, $P$ is the gas pressure, and $\rho$ is the mass density.
Unless otherwise stated, all thermodynamic variables are understood to be
$\phi$ independent azimuthal averages.  The velocity $\bb{v}$
will in general contain (convective) fluctuating components; any azimuthal
averaging will be treated explicitly.  The SCZ gravitational
field $g$ is to a good approximation $GM_\odot/r^2$, where $GM_\odot$
is the product of the Newtonian gravitational
constant and a solar mass.    Finally, we define a dimensionless entropy function
$\sigma$:
\beq
\sigma \equiv
\ln P\rho^{-\gamma},
\eeq
where $\gamma$ is the adiabatic index.  The thermodynamic entropy is then given
by $S\equiv C_P\sigma$, where $C_P$ is the specific heat at constant pressure.  

\subsubsection {Review of the TWE solution}

Our starting point is the partial differential equation
arising from the assumption of
a dominant thermal wind balance in the time averaged
(and thus azimuthally averaged) vorticity
equation (e.g. Kitchatinov \& R\"udiger 1995, Thompson et al. 2003,
Miesch 2005, B09):
\beq\label{twe}
R {\dd\Omega^2\over \dd z} 
={g\over \gamma r } {\dd\sigma \over  \dd\theta}
\eeq
If, for some reason, there is a functional relationship of the 
form $\sigma=f(\Omega^2)$ (the sign of $\Omega$ presumably does
not matter), then the TWE may
be written in spherical coordinates as
\beq\label{maineq}
{\dd \Omega^2\over \dd r}
-
\left(
{g f' \over \gamma r^2\sin\theta \,\cos\theta}+
{\tan\theta\over r}
\right)
{\dd \Omega^2 \over\dd\theta} =0,
\eeq
where $f'=df/d\Omega^2$.  Notice, however, that exactly the same equation 
obtains if the functional relationship between $\sigma$ and $\Omega$
takes the more general form
\beq\label{gauge}
\sigma'\equiv \sigma-\sigma_r=f(\Omega^2),
\eeq
where $\sigma'$ will be referred to as residual entropy,
and $\sigma_r$ is any function of $r$ alone.  We will make important use
or this ``gauge freedom'' later in this paper.

The solution of equation (\ref{maineq}) is that $\Omega^2$ is constant
along the characteristic contours (B09):
\beq
R^2=r^2\sin^2\theta = A -{B\over r},
\eeq
where $A$ is a constant of integration, and 
\beq
B= - {2GM_\odot f'(\Omega^2)\over\gamma}.
\eeq
If we denote by the subscript `0' the fiducial starting point
of the characteristic (at which $\Omega$ is specified), then
the contour takes the form
\beq\label{chars}
R^2=r^2\sin^2\theta =  r_0^2 \sin^2\theta_0- B\left(  {1\over r}-
{1\over r_0} \right)
\eeq

While $B$ is a constant along a given characteristic, its value can
change from one characteristic to another depending upon the form
of the function $f'$.  B09 showed, however, that the simplest
parameterizations of $f'$ ({\em e.g.,} $f'=$ a global constant or $f'=$ a linear
function of $\sin^2\theta_0$) already 
give excellent qualitative fits to the
helioseismology data.  In fact, it is not difficult to do even better. 

Figure [1] shows a comparison between isorotational contours (in
black) from recent GONG data\footnote{We thank R. Howe for providing
us with these beautiful results.} and the characteristics of the
thermal wind equation (overlayed in white).  Here, we have carried
out a more systematic approach than that of B09: we have integrated
each characteristic from an {\em interior} starting point of $0.9$
solar radii (as opposed to the less accurate surface fitting in B09),
and chosen the constant $B$ to match the initial slope of the data at
the fitting radius.  The new result is remarkably accurate, not just in
qualitative form, but in quantitative detail as well.  Expected departures
of the black and white contours in the surface layers and tachocline
are plainly visible, yet in the bulk of the SCZ the match is excellent.

Note the equator and poles, where the curvature of the contours is,
perhaps surprisingly, reproduced with great precision.\footnote{The
inversion process by which the rotation contours are determined from
helioseismology data is, it should be noted, less reliable near the
poles.}  While this {\em is} evidence of the validity of thermal
wind balance, the implied functional relationship between $\Omega$
and $\sigma$ (or $\sigma'$) is also at least partly a consequence
of mathematical symmetry.  In the polar neighborhoods $R$ gradients
are small, and near the equator $z$ gradients are small, so that in
both regions azimuthally averaged flow quantities will be a function
predominantly of spherical $r$.  These quantities must therefore be
functionally related.  This functional relationship evidently extends
seamlessly into the regions immediately adjacent to the symmetry points,
in which the isocontours are not predominantly radial.

\begin{figure}
\epsfig{file=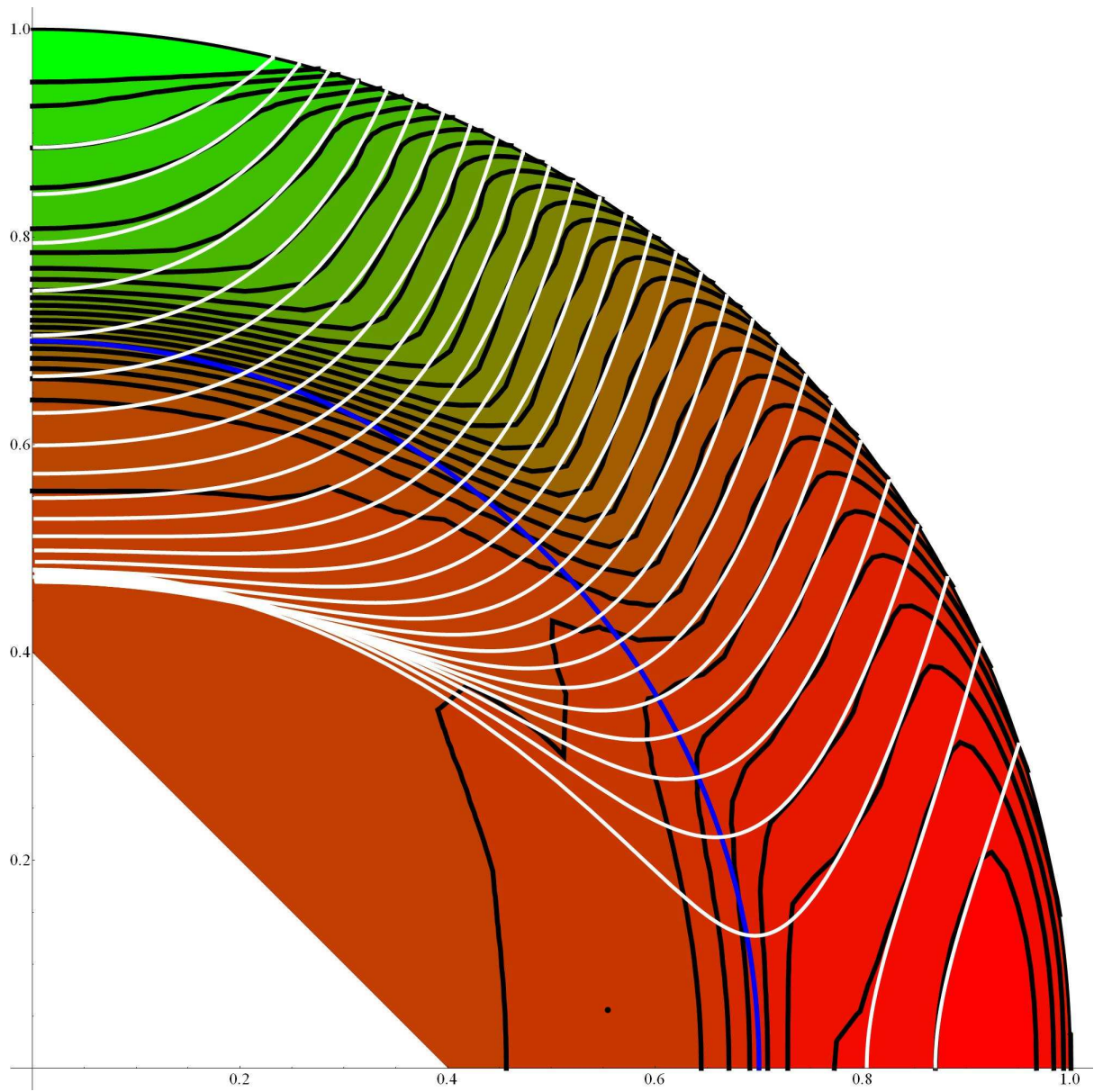, width=16cm}
\caption {Constant $\Omega$ contours (\ref{chars})
of the thermal wind equation
(\ref{twe}) (white
curves) plotted on top of (black) isorotation contours from helioseismology
data (GONG results courtesy of R. Howe).  Blue contour
is the bottom edge of the convective zone.  Scale is in solar radii.
Away from the tachocline and outer surface layers, the match is
excellent.  See text for further details.}
\end{figure}

\subsection {A functional relation between
residual entropy and angular velocity}

\subsubsection {Theoretical considerations}

One possible explanation for the striking resemblance of the analytic
isorotation contours to the solar data was advanced in B09 by invoking
a weak magnetic field.  In the presence of such a field, disturbances
associated with magnetobaroclinic modes lead to a condition of marginal
instability if $\sigma=f(\Omega^2)$.  This is a {\em dynamical} connection
between $\sigma$ and $\Omega$, indeed an MHD connection.

We will review the merits and shortcomings of this scenario in more
detail in \S 4.   Here, we explore a more general connection between
$\sigma$ and $\Omega$ relying neither upon magnetic fields, nor even
upon dynamical constraints.  The idea relies instead on the gauge freedom
expressed in equation (\ref{gauge}).

Consider a fiducial, nonrotating, spherical, convecting star that
maintains a well-defined, long-term average radial entropy profile
$\sigma_r$.  Convection continously mixes high and low entropy fluid
elements, yet $\sigma_r$ remains fixed.  The reason for this is not
mysterious.  The system is heated from below, and this input tries to
drive the gas into a state with a considerably steeper entropy profile.
Convection counters this, allowing a balance to be struck between the
external thermal driving and the heat transport by turbulent fluid
elements.  In particular, convective displacements do not need to be
within constant entropy surfaces to preserve the long-term $\sigma_r$
profile.  This profile is already being maintained by the external
heating.

If we now introduce a small amount of rotation into the problem, we
expect Coriolis forces to produce a more complex, but still well-defined,
long-term entropy profile, $\sigma(r,\theta)$.  As before, convection
will try its very best to change this profile.   However, in contrast to
the spherical case of the previous paragraph, it is very far from clear that
radial heating from below can maintain a steady, nonspherical entropy profile.
On the other hand, if the convective velocities $\bb{v}$ on average
satisfy the constraint
\beq\label{conconst}
\bb{v \cdot\nabla}\sigma'  =0
\eeq 
then the entropy profile certainly {\em can} be maintained.  In this case, the
long term maintenance
of the spherically averaged component of the entropy $\sigma_r$ is assured by 
heating from below, while the residual entropy $\sigma'$
is preserved because convective displacements will tend
on average to occur in surfaces
in which this quantity is constant.  Thus, the long-term average entropy
profile $\sigma(r,\theta)$ remains intact.  Convecting elements move not
in constant entropy surfaces (which would preclude heat transport!),
but in surfaces of constant {\em residual} entropy $\sigma'$.

Some care is needed for the proper interpretation of equation
(\ref{conconst}).  The poloidal components of $\bb{v}$ (hereafter
$\bb{v_p}$) have a long term average value much smaller than a typical
fluctuation amplitude $|\bb{v_p}|$.  (This systematic velocity manifests
itself principally as the meridional flow that has been measured at,
and near, the surface.)  A time average of equation (\ref{conconst})
would produce contributions of second order in the fluctuations, including
heretofore neglected $\sigma'$ fluctuations.  On the other hand, the fact
that the ``up-down'' convective velocity fluctuations nearly cancel means
that for a given convective cell, there is very little to distinguish
these two directions.  (Because of rotation, ``upward'' and ``downward''
need not be precisely radial, of course.)  These first order velocity
fluctuations in essence establish a well-defined axis relative to which
$\del\sigma'$ is orthogonal; the positive and negative sense of the axis
is unimportant.  A more precise formulation of equation (\ref{conconst})
is therefore
\beq \langle \left( \bb{v \cdot\nabla}\sigma'\right)^2
\rangle \ll \langle |\bb{v_p}^2| \rangle \langle |\del\sigma'|^2\rangle
\eeq
where the angle brackets $\langle\ \rangle$ denote time averaging.
This is also a more robust formulation, since the quantity on the left
is now dominated by the first order velocity contribution $\bb{v_p}$,
and fluctuations in $\sigma'$ do not enter.

More physically, simulations and laboratory experiments suggest that
convective transport is dominated by long-lived coherent structures.
The updrafts and downward plumes characteristic of these structures
are clearly associated with dominant ``first order'' velocity terms,
and it is these structures that would exist within and maintain constant
$\sigma'$ surfaces.  If it could now be further argued that convecting
fluid elements (or coherent structures) also lie within constant
$\Omega$ surfaces, we would arrive at an understanding of why there
is a functional relation between $\Omega$ and $\sigma'$: if elements
are simultaneously moving in $r\theta$ surfaces of constant $\Omega$
and $\sigma'$, these must be the same surfaces.  In other words, the
two quantities are functionally related.  In the next section, we will
see that there is indeed a simple argument leading to the conclusion
that displaced fluid elements will inevitably move in constant $\Omega$
surfaces.

Finally, we reiterate that although the original B09 derivation of the
isorotation contours was based on the assumption that $\sigma=f(\Omega^2)$,
exactly the same contours emerge if the functional relation is
$\sigma'=f(\Omega^2)$.  This very simple gauge invariance of the thermal
wind equation allows for a considerably wider, more general class of
thermal wind solutions, eliminating the constraint that constant (total)
entropy and angular velocity surfaces coincide.

\subsubsection {Comparison with simulations}

Support for the line of argument advanced in the previous section can be
found in the (purely hydrodynamical) MBT06 simulations.  In these runs,
the residual entropy $\sigma'$ is calculated relative to a background
radial profile, corresponding to our $\sigma_r$.  In figure [2], we
show time averaged contours of the angular velocity profile $\Omega$
from the MBT06 run designated AB3, of the residual entropy, and of the
total entropy $\sigma$.  The agreement between the $\Omega$ and $\sigma'$
contours is apparent, as is the gross disagreement between $\Omega$ and
$\sigma$.  Evidently, there is a functional relation between $\sigma'$ and
$\Omega$, whereas there is no such relation between $\Omega$ and $\sigma$.
This is in good agreement with the above discussion.
Notice as well the small gradient of $\sigma'$ near the equator.
This is as expected if $\sigma_r$ represents an angular average
of $\sigma$, which would be dominated by $\theta$ values near $\pi/2$.

\begin{figure}
\epsfig{file=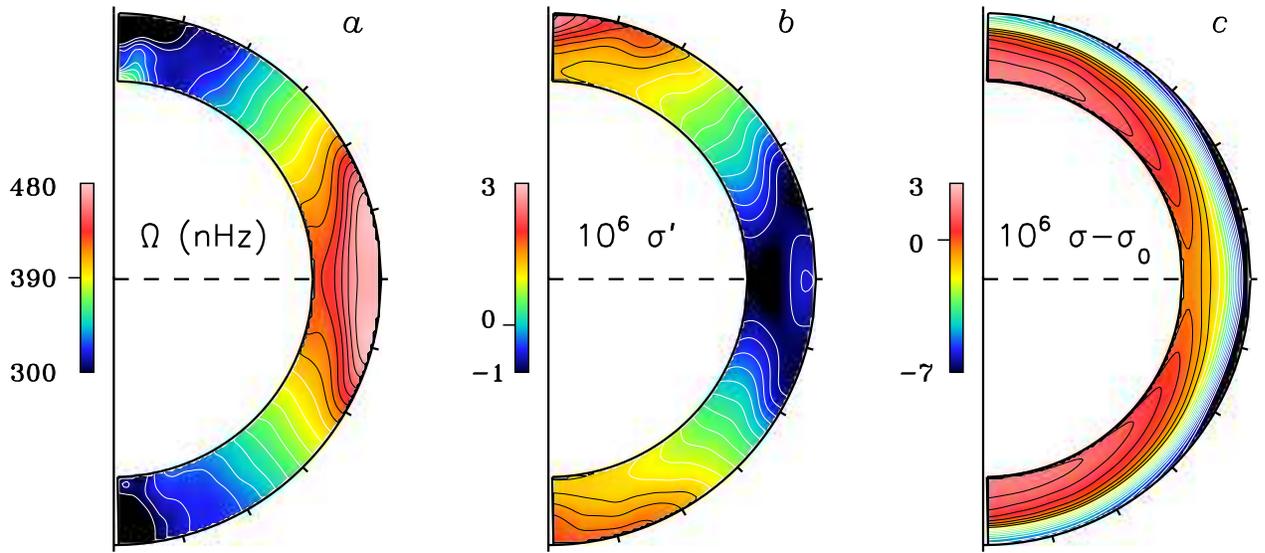}
\caption{Isorotation contours (a), contours of constant residual entropy
(b), and contours of total entropy (c), from the run AB3 of MBT06.
(In [c], a constant background entropy has been subtracted out.)
There is generally very good agreement between the angular velocity
and residual entropy contours, but not between the angular velocity and
total entropy contours.  (Figures courtesy of M. Miesch.)} \end{figure}

\section {Nonaxisymmetric disturbances in a differentially
rotating medium} 

\subsection {Comoving coordinates and wavenumbers}

Consider the behavior of convective disturbances in a differentially
rotating gas $\Omega(R,z)$.  The disturbances are embedded in this
shearing medium, and since they are not propagated as waves, we may
think of them as local in character.  But ``local'' refers to the
shearing gas, not to an absolute space-time grid, so it is particularly
useful to express flow quantities in Lagrangian coordinates tied to
the shear flow.
We designate these coordinates as the ``primed'' system.
This transformation is given by
\beq
R'=R, \quad \phi' = \phi -\Omega(R,z) t, \quad z'=z, \quad t'=t.
\eeq

For our present purposes, we need consider only the 
equation of mass conservation: 
\beq
\del\bcdot(\rho \bb{v})=0,
\eeq
where $\bb{v}$ is understood to be the velocity relative
to the differential rotation $R\Omega$.  
The gradients are of course the standard Eulerian derivatives
taken at fixed time.   We have not written the explicit $\rho$ time
derivative, since we are ultimately interested in long term time averaging
and the term will disappear.
Apart from this, the equation is exact; the mass flux is taken to have 
a full space time dependence.  

Transforming to our comoving system, 
\beq
{\dd\ \over \dd R}= {\dd\ \over \dd R'} -t {\dd\Omega\over \dd R}
{\dd\ \over\dd\phi'}, 
\eeq
\beq
{\dd\ \over \dd z }= {\dd\ \over \dd z'} -t {\dd\Omega\over \dd z}
{\dd\ \over\dd\phi'},  
\eeq
and the $\phi$ derivative is unchanged.
More compactly,
\beq
\del = \del' -(t\del\Omega)
{\dd\ \over\dd\phi'},
\eeq
a relation that holds for all three components.  

The embedded mass flux field $\rho \bb{v}$ may be expressed as a superposition
of Fourier components of the form
\beq
\exp\left[i\left(k'_R R'  +m\phi'+k'_z z'-\omega t' \right)\right]
\eeq
in which all components of the $\bb{k'}$ wave vector are constants.
The above coordinate transformation implies that in the mass conservation equation,
the standard Eulerian spatial derivatives with respect to $R$
and $z$ are replaced respectively by $ik_R(t)$ and $ik_z(t)$, where
\beq
k_R(t) = k'_R -mt{\dd\Omega\over \dd R},
\eeq
\beq
k_z(t) = k'_z -mt{\dd\Omega\over \dd z}
\eeq
Henceforth, the time dependence of $k_R$ and $k_z$ will be understood.
A somewhat similar formalism has been developed for embedded disturbances in shear
turbulence, where it is known as {\em rapid distortion theory} (Townsend 1980).

\subsection{Mass flux morphology}                         

The Fourier series for the mass flux is:
\beq
\rho\bb{v} = \sum  \bb{\mu} (\bb{k'}, \omega)
\exp\left[i\left(k'_R r'  +m\phi'+k'_z z'-\omega t' \right)\right],
\eeq
where $\bb{\mu} (\bb{k'}, \omega)$ is the amplitude of the indicated wavenumber
and frequency.  Mass conservation leads immediately to
\beq
0=\del\bcdot(\rho\bb{v}) = 
 \sum  \bb{k}\bcdot \bb{\mu} (\bb{k'}, \omega)
\exp\left[i\left(k'_R r'  +m\phi'+k'_z z'-\omega t' \right)\right].
\eeq
Each Fourier term is linearly independent of the others, hence for all
$\bb{k'}$, $\omega$,
\beq
(\bb{k'}-mt\del\Omega)\bcdot \bb{\mu} (\bb{k'}, \omega)=0
\eeq
For nonvanishing $m$, at sufficiently large $t$,
\beq
\bb{\mu}(\bb{k'}, \omega)\bcdot 
\del\Omega = 0.
\eeq
The mass flux is a superposition of the $\bb{\mu}$ vectors.  Therefore,
if the flux is not dominated by an $m=0$ component, it will tend toward
surfaces of constant $\Omega$.  Note that this statement is restricted
neither to linearized displacements, nor to the WKB limit.  Finally, it
has been noted that this tendency for turbulent convective structures to
align with constant $\Omega$ surfaces may also be supported on a dynamical
basis via the effects of the Coriolis force on plumes (Brummell, Hurlburt,
\& Toomre 1996).

Alignment of embedded structures with constant $\Omega$ surfaces is of
course seen regularly in accretion disk simulations, where the shear is
large and depends only upon $R$.  Disks features often appear nearly
axisymmetric because of the effect of the shear.  In the Sun, the $R$
and $z$ dependence of $\Omega$ leads to a more structured response.
This assumes, of course, that the rotational shear is strong enough to
interact with the convective cells over the course of their lifetime.  
This is a reasonable assumption for the Sun, especially if the convection
is dominated by long-lived coherent stuctures.  

This, in principle, supplies the missing link of the argument of \S 2.2.1:
the course of a displaced mass element will tend on average to follow
a constant $\Omega$ surface, essentially because differential rotation
wraps the flow into sheets of constant $\Omega$.  Since fluid elements
move at once in surfaces of constant $\Omega$ and constant $\sigma'$,
these surfaces must in general coincide\footnote{The degenerate case in
which the surfaces do not coincide would correspond to trivial circular
azimuthal orbits of the fluid element.}.

Notice how the complete set of three hydrodynamical equations ---
mass conservation, entropy conservation, and the equation of motion ---
has been incorporated in our solution.  From mass conservation,
we infer that fluid elements move in constant $\Omega$ surfaces.
From entropy conservation, we infer that fluid elements move in surfaces
of constant $\sigma'$.  From these two inferences, it follows that
constant $\Omega$ and constant $\sigma'$ surfaces coincide, and this,
expressed as a functional dependence, is used in the equation of motion
(thermal wind balance) to construct the isorotational contours of figure
[1].  

\subsection {Heat transport}

The convective heat transport is 
\beq
\bb{Q} \equiv \rho\bb{v}\delta w,
\eeq
where $\delta w$ represents the difference between the specific enthalpy
of a convecting element and its surroundings.  The reasoning we used
in the last section shows that 
\beq\label{QQ}
\bb{Q\cdot\del}\Omega \rightarrow 0,
\eeq
as time progresses.  In other words, the heat flux, like the mass flux,
is predominantly in surfaces of constant $\Omega$.

This offers a very plausible physical basis for what has long been
regarded as a puzzling issue:  why are the solar rotation contours
quasi-radial?  The answer we suggest is that the rotation contours are
also the natural conduits for convective heat transport, and their
near radial character simply reflects this property.  The contours
represent a compromise between strong radial thermal driving and the
dynamical exigencies embodied in the TWE.  Thus, they are nearly,
but not perfectly, radial throughout the bulk of the SCZ.  Near the
equator and poles, the convective heat flow certainly does {\em not}
follow isorotation contours, but as has already been noted in \S 2.1.1,
these are precisely the regions in which the functional relationship
between $\sigma'$ and $\Omega$ is maintained by mathematical symmetry.

It is interesting to speculate on how the isorotation contours would
change in rapidly rotating stars.  For example, when a solar-type star
first arrives on the main sequence it spins much more rapidly than
the Sun does now.  The convection pattern is likely to be dominated
by Coriolis effects, expressed through the Taylor-Proudman constraint
(Busse \& Simitev 2007).  Numerical models confirm that convection
takes the form of elongated ``banana cells'' at low latitudes, outside
the tangent cylinder that encloses the radiative zone (Ballot, Brun \&
Turck-Chi\`eze 2007; Brown et al. 2008).  In these rapidly rotating stars,
the isotachs tend to be roughly cylindrical and there are pole-equator
differences in entropy and temperature that are significantly larger
than those inferred for the Sun.  In our model, the $1/\Omega^2$ scaling
of $f'(\Omega^2)$ suggests a much smaller value for the $B$ parameter in
rapid rotators, which does indeed lead to more cylindrical isotachs.

\section {Concluding discussion}

What is novel about our approach is not the use of the thermal wind
equation {\it per se,} which has provided important guidance to
workers in this field for many
years now.  It is the suggestion that there is a functional relationship
between the entropy and angular velocity, for it is this that allows
an analytic deduction of the shape of the Sun's isorotation contours
to be made directly from the TWE.

As a formal matter, there is some freedom in choosing the precise
form of the entropy--angular velocity relation, a type of
gauge invariance.  This is because the TWE involves only the $\theta$
derivative of the entropy and the $z$ derivative of the angular velocity.
Different physical mechanisms for coupling $\sigma$ and $\Omega$ in
principle entail a different choice of gauge.

In B09, an MHD process was invoked to couple $\sigma$ and $\Omega$.
That paper noted that baroclinic modes in a weakly magnetized gas become
marginally unstable if constant $\sigma$ and constant $\Omega$ surfaces
coincide, i.e. $\sigma = f(\Omega^2)$.  In that approach, rotation and
stratification are set on an equal footing in determining the dynamical
stability of the Sun's outer layers.   
Moreover, in the MHD model, the radial entropy gradient is restricted to
be significantly less than the $\theta$ gradient.  This can be readily
deduced from figure [1], since in the $\sigma=f(\Omega^2)$ interpretation,
the GONG data curves are isentropes as well as isorotation contours.
Whether the actual radial entropy gradient is constrained in this way
depends upon the scale of the effective mixing length.  If this length
is only a small fraction of the radial extent of the convection zone, the
radial entropy gradient could significantly exceed the $\theta$ gradient.

Finally, in the MHD model, the goodness of the fit between the theoretical
and observed isorotation contours depends upon a very tight functional
relationship between $\sigma$ and $\Omega$.  But were the $\sigma$
and $\Omega$ surfaces to coincide ${\it exactly}$, convection would
be completely cut-off.  Conversely, as we have noted above, if very
vigorous convection is required, it becomes untenable to maintain a
$\sigma=f(\Omega^2)$ relation.  Is the departure between
$\del\sigma$ and $\del\Omega$ large enough to allow for vigorous
convection, yet small enough to ensure that $\sigma=f(\Omega^2)$        
remains an excellent approximation in the TWE?

The purely hydrodynamical explanation for the solar isorotation contours
avoids many of these potential difficulties.  The restriction on the
size of the radial entropy gradient is lifted by seeking a functional
relationship of the form of equation (\ref{gauge}).  As the residual
entropy $\sigma'$ can be significantly smaller than $\sigma$,  the
effects of a potentially large radial entropy gradient can be eliminated
from the rotational dynamics of the problem.  The physical basis for
this functional form is related to the fact that convection should
occur on average in surfaces of constant $\sigma'$, so as to maintain
a well-defined time averaged $\sigma$ (total entropy) profile.
The kinematics of differential rotation and simple mass conservation
ultimately confine mass motions to surfaces of constant $\Omega$.
Together with equation (\ref{conconst}), this ensures a relationship
of the form $\sigma'=f(\Omega^2)$, and even suggests why the contours
are quasi-radial (cf. eq. [\ref{QQ}]).  Without the need to restrict
$\dd\sigma/\dd r$, convection can be as vigorous as needed
(provided that it does not dominate the differential rotation), yet not
directly affect the TWE solution for the isorotation contours.

Finally, the ``hydrodynamical'' argument can easily accommodate
MHD processes.  Neither mass conservation nor entropy conservation,
the two physical processes central to our argument, change in the 
presence of a magnetic field.  The dynamical details, in particular
the rotational couplings, may well be altered, but the functional
relation between $\sigma'$ and $\Omega$ should remain robust.  
We therefore suggest that the simple physical arguments
leading to equation (\ref{gauge}) may prove to be a suitable
basis for understanding much of the dynamics that gives
rise to the solar isorotation contours.  

The theory outlined in this paper is meant to be a viable alternative
to an MHD model, and it is as yet far from complete.  
Our presentation falls well short of the standards of mathematical proof.
We have relied on ``tendencies'' for the fluid to move in surfaces
of constant $\Omega$ and $\sigma'$ to be strong enough to firmly
establish the relationship expressed compactly by equation (\ref{gauge}).
But figure [1] speaks for itself.  To understand at a deeper level why
this approach seems to work so well, a more detailed elucidation of the
dynamics is clearly desirable.

To conclude, in this paper we have advanced theoretical arguments to
explain the striking correspondence between isentropes and isotachs in
azimuthally averaged models of the solar convection zone.  These arguments
need to be backed up by directed and detailed computations.  To this
end, we are currently developing an anelastic code that can be used to
demonstrate explicitly that the relationship of equation (\ref{gauge})
is valid for a variety of relevant configurations.  Our principal
aim will be to explore the connection between $\Omega$ and $\sigma$
(or $\sigma'$) as the overall rotation rate is varied.  The effects on
this relationship of varying the thermal boundary conditions, in order to
represent the influence of the solar tachocline (Miesch \& Toomre 2009),
are also of great interest.

\section*{Acknowledgments}
We thank the referee M. Miesch for a very stimulating exchange that
greatly sharpened our understanding of the numerical simulations as well
as many of the issues presented here, and for kindly providing us with
figure [2] of this paper.  SAB would also like to thank J. Goodman and P.
Lesaffre for their insightful comments.  We are grateful to R. Howe for
generously making her reduced GONG data available to us.  NOW expresses
his gratitude to the Laboratoire de Radioastronomie at the Ecole Normale
Sup\'erieure, the Institut de Physique du Globe of the Universit\'e de
Paris 7, and the Universit\'e de Paris 6 for their generous financial
support during the time that this work was completed.  This work has
been supported by a grant from the Conseil R\'egional de l'Ile de France.


\begin{thebibliography}{}
\bibitem[]{b09}Balbus S. A. 2009, MNRAS, 395, 2056 (B09)
\bibitem[]{bht96} Brummell, N. H., Hurlburt, N. E., \& Toomre, J. 1996,
ApJ, 473, 494
\bibitem[]{bbt07}Ballot, J., Brun, A. S., \& Turck-Chi\`eze, S. 2007, ApJ, 669, 1190
\bibitem[]{bbbmt08}Brown, B. P., Browning, M. K., Brun, A. S., Miesch, M. S., \& Toomre, J. 2008,
ApJ, 689, 1354
\bibitem[]{bs07} Busse, F. H., \& Simitev, R. D. 2007,  in ``Mathematical Aspects of Natural Dynamos'', 
eds. E. Dormy \& A. M. Soward (Boca Raton: CRC Press), p. 168 
\bibitem[]{kr95}Kitchatinov, L. \& R\"udiger, G. 1995, A\&A, 299, 446        
\bibitem[]{m05}Miesch, M. S. 2005 Living Revs. Sol. Phys., 2, 1.
(www.livingreviews.org/lrsp-2005-1)
\bibitem[]{mbt06}Miesch, M. S., Brun, A. S., Toomre, J. 2006, ApJ, 641, 618
(MBT06)
\bibitem[]{mbdt08} Miesch, M. S., Brun, A. S., DeRosa, M. L., \& Toomre, J. 2008,
ApJ, 673, 557
\bibitem[]{mt09}Miesch, M. S., \& Toomre, J. 2009, ARA\&A, 41, 317 
\bibitem[]{tcmt03} Thompson, M. J., Christensen-Dalsgaard, J., Miesch, M. S.,
\& Toomre, J. 2003, ARAA, 41, 599
\bibitem[]{t80} Townsend, A. A. 1980, The Structure of Turbulent Shear Flow
(Cambridge: Cambridge University), p. 43.
\end{thebibliography}
\end{document}